\documentclass[conference]{IEEEtran}
\IEEEoverridecommandlockouts
\IEEEoverridecommandlockouts
\usepackage{cite}
\usepackage{amsmath,amssymb,amsfonts}
\usepackage{graphicx}
\usepackage{textcomp}
\usepackage{balance}
\usepackage{xcolor}
\usepackage{algorithm}
\usepackage[noend]{algpseudocode}
\usepackage{mathtools}
\usepackage{multirow}
\usepackage{enumitem}
\usepackage{makecell}
\usepackage{subcaption}
\usepackage{calc}
\usepackage{comment}
\usepackage{etoolbox}
\usepackage{booktabs}
\usepackage{adjustbox}
\usepackage{listings}
\usepackage{xcolor}

\usepackage{url}

\usepackage[breaklinks]{hyperref}
\hypersetup{colorlinks,
}
\usepackage{siunitx}
\DeclareSIUnit{\coeq}{CO_2e}
\usepackage{tikz,pgfplots,pgfplotstable}
\pgfplotsset{compat=1.9} 
\newcommand{\mystar}{{\fontfamily{lmr}\selectfont$\star$}}

\definecolor{codegreen}{rgb}{0,0.6,0}
\definecolor{codegray}{rgb}{0.5,0.5,0.5}
\definecolor{codepurple}{rgb}{0.58,0,0.82}
\definecolor{backcolour}{rgb}{0.95,0.95,0.92}

\lstdefinestyle{mystyle}{
    backgroundcolor=\color{backcolour},   
    commentstyle=\color{codegreen},
    keywordstyle=\color{magenta},
    numberstyle=\tiny\color{codegray},
    stringstyle=\color{codepurple},
    basicstyle=\ttfamily\footnotesize,
    breakatwhitespace=false,         
    breaklines=true,                 
    captionpos=b,                    
    keepspaces=true,                 
    numbers=left,                    
    numbersep=5pt,                  
    showspaces=false,                
    showstringspaces=false,
    showtabs=false,                  
    tabsize=2,
    moredelim=**[is][\color{red}]{@}{@}1
}
\def\BibTeX{{\rm B\kern-.05em{\sc i\kern-.025em b}\kern-.08em
    T\kern-.1667em\lower.7ex\hbox{E}\kern-.125emX}}
    
\begin{document}

\title{
SCARIF: Towards Carbon Modeling 
of \\ Cloud Servers with Accelerators\\

}

\author{
Shixin Ji\IEEEauthorrefmark{1},
Zhuoping Yang\IEEEauthorrefmark{1},
Xingzhen Chen\IEEEauthorrefmark{1},
Stephen Cahoon\IEEEauthorrefmark{1},
\\
Jingtong Hu\IEEEauthorrefmark{1},
Yiyu Shi\IEEEauthorrefmark{2},
Alex K. Jones\IEEEauthorrefmark{1},
Peipei Zhou\IEEEauthorrefmark{1}
\\
\IEEEauthorrefmark{1} University of Pittsburgh,
\IEEEauthorrefmark{2} University of Notre Dame
\\
Email:{\tt \{shixin.ji, peipei.zhou\}@pitt.edu
}
}

\maketitle

\begin{abstract}
Embodied carbon has been widely reported as a significant component in the full system lifecycle of various computing systems green house gas emissions.  Many efforts have been undertaken to quantify the elements that comprise this embodied carbon, from tools that evaluate semiconductor manufacturing to those that can quantify different elements of the computing system from commercial and academic sources.  However, these tools cannot easily reproduce results reported by server vendors' product carbon reports and the accuracy can vary substantially due to various assumptions.  
Furthermore, attempts to determine green house gas contributions using bottom-up methodologies often do not agree with system-level studies and are hard to rectify.  
Nonetheless, given there is a need to consider all contributions to green house gas emissions in datacenters, we propose SCARIF, the \underline{S}erver \underline{C}arbon including \underline{A}ccelerator \underline{R}eporter with \underline{I}ntelligence-based \underline{F}ormulation tool.  
SCARIF has three main contributions: 
(1) We first collect reported carbon cost data from server vendors and design statistic models to predict the embodied carbon cost so that users can get the embodied carbon cost for their server configurations. 
(2) We provide embodied carbon cost if users configure servers with accelerators including GPUs, and FPGAs. 
(3) By using case studies, we show that certain design choices of data center management might flip by the insight and observation from using SCARIF. 
Thus, SCARIF provides an opportunity for large-scale datacenter and hyperscaler design. 
We release SCARIF as an open-source tool at \textbf{\emph{\url{https://github.com/arc-research-lab/SCARIF}}}.
\end{abstract}

\vspace{-10pt}
\section{Introduction}
Datacenters have become a significant source of energy consumption globally.  
Although they currently consume approximately 1\% of energy consumption worldwide~\cite{5ways}, estimates have them reaching or exceeding 10\% globally in the next five to ten years~\cite{Electricity}.
This has raised concerns about the greenhouse gas (GHG) resulting from powering these datacenters.

Until recently, many discussions have been made about the GHG in the manufacturing phase i.e. the embodied carbon cost.
Several tools have been proposed to measure the embodied GHG emissions of chips and systems~\cite{igsc2016greenchip,KLINE2019322,ACT}.  
Unfortunately, embodied carbon costs for computing systems are hard to measure.
In other sectors, such as civil engineering, embodied GHG emissions can be measured by studying the materials used 
such as concrete or steel.  
However, semiconductor manufacturing cannot use the same process as the bulk of the materials used in the manufacturing process are temporary or etched away through the process.  Moreover, to protect intellectual property, fabrication techniques are closely guarded secrets.  
Thus, only a limited number of studies exist for estimating the embodied 
GHG emissions from semiconductor manufacturing~\cite{imec-dtco20,Boyd-LCA-Semi-12}, which may, or may not be representative compared to production semiconductor techniques.  
Thus, determining embodied GHG emissions using a bottom-up strategy has led to significant inaccuracies~\cite{ACT}.  
However, for sustainable computing research to proceed, tools that can provide estimates of systems deployed in data centers are required. 
Moreover, these tools must be able to reasonably extrapolate the cost of including custom accelerator hardware such as graphics processing units (GPUs), tensor processing units (TPUs), and field programmable gate arrays (FPGAs).  
While it is difficult to validate the exact results of these estimates to the \SI{}{\kilo\gram\coeq} (carbon dioxide equivalent), tools that allow exploration of different relative design choices while considering the whole system
embodied GHGs are needed.

\begin{figure}[tb]
\centering
\includegraphics[width=1\columnwidth]{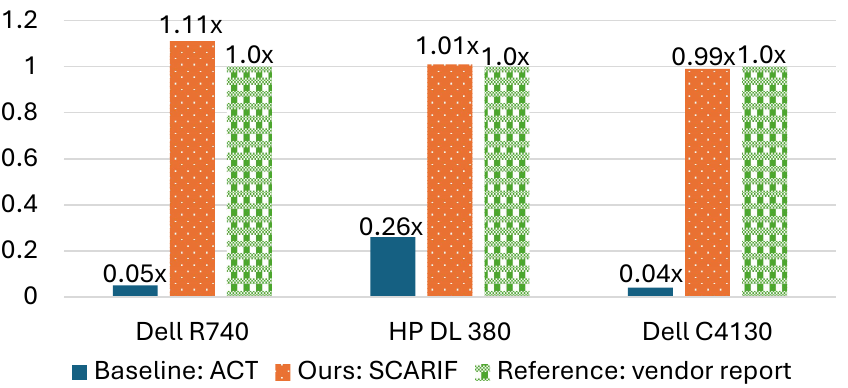}
\vspace{-15pt}
\caption{The comparison of embodied carbon estimation among  ACT, SCARIF (ours), and the vendor reports on different server settings.
The carbon costs are normalized to the vendor reports. 
SCARIF (ours) significantly improves the accuracy.}
\vspace{-20pt}
\label{fig1}
\end{figure}

\textit{Towards this end, we propose SCARIF, or the \underline{S}erver \underline{C}arbon including \underline{A}ccelerator \underline{R}eporter with \underline{I}ntelligence-based \underline{F}ormulation tool.}  SCARIF builds a server model for GHG emissions of a generic data center server based on simple configuration parameters 
which are extracted from carbon footprint reports of various servers.
As shown in Figure~\ref{fig1}, we demonstrate that our model can predict the results with better accuracy for first-order estimations compared with prior tools. 
We then demonstrate how these servers can be estimated to be equipped with accelerators based on data combined with estimates from the ACT tool~\cite{ACT} normalized to system estimates from the SCARIF model. 
SCARIF can leverage other tools for measuring accelerator details or even direct measurements, allowing detailed holistic comparisons of GHG emissions from different system candidates.

In summary, our main contributions are:

\begin{itemize}
\item We first summarize and analyze the current state-of-the-art server carbon emissions report from different vendors including HP, Dell, and Lenovo (Section~\ref{sec:observations}). 
\item We present the first tool, SCARIF,
to estimate the server carbon emissions based on high-level criteria of system configuration within the accuracy threshold of existing full-product lifecycle studies of server products from multiple vendors (Section~\ref{sec:model_build} \& \ref{sec:model_valid}). 
\item We use SCARIF
to estimate and analyze the total carbon emissions of
the modeled server with different accelerator hardware, including GPU and FPGA (Section~\ref{sec:model_accelerator}).
\item 
We perform case studies to 
discuss how to use SCARIF to guide the upgrading policy in the data center and how that impacts the design choices (Section~\ref{sec:analysis}).
\end{itemize}
\vspace{-5pt}
\section{Related Works}
\label{sec:related}
\vspace{-5pt}
The growing awareness of sustainability and carbon footprints has significantly impacted various aspects of society. 
hardware vendors like Apple~\cite{Apple-ENV}, Google~\cite{Google-ENV}, Dell~\cite{Dell-ENV}, HP~\cite{HP-ENV}, and Lenovo~\cite{Lenovo-ENV}, among others, have begun to report the carbon footprints of their products. Upstream in the information and communication technologies supply chain, semiconductor manufacturers such as TSMC~\cite{TSMC_report} and SK Hynix~\cite{SK_Hynix_report} also disclose their reports of corresponding technology nodes while other vendors such as Intel and Samsung report more aggregated data. 

In the academic community, research efforts have made considerable progress over the last decade to estimate or profile the hardware systems' carbon cost and use the profiling information to guide the design toward sustainable systems.
In recent years, life Cycle Assessment (LCA) has become a common method to quantitatively evaluate the GHG emissions of computing systems throughout their whole lifetime. 
PAIA\cite{PAIA_tool}is a commercial and closed-source LCA tool which is widely used by the hardware vendors like Levono, HP, and Dell.
Other than the LCA tools, more carbon cost estimation and analysis approaches are proposed to derive insights into designing sustainable computing systems. Greenchip~\cite{igsc2016greenchip,KLINE2019322} is the earliest predictive estimation tool to comprehensively understand the environmental impact of computing systems.
More recently, ACT~\cite{ACT} is a system modeling tool that is built, like Greenchip, with lower-level data from industry fabs' but with a focus on massage system level parameters to better calibrate to technology companies' LCA reports and can estimate systems with different types of system-on-chip (SoC) and storage devices. 
Moreover, based on these carbon cost estimation tools, more studies are conducted to gain insights into more specific domains~\cite{li2023toward,eeckhout2024focal,faiz2023llmcarbon,kim2023greenscale,zhou2023refresh,yang2024chiplet,zhuang2024ssr,zhuang2023dac,yang2023aim}.
\vspace{-5 pt}
\section{Observations}
\vspace{-5pt}
\label{sec:observations}
In this section, we analyze the hardware vendors' reports in Section~\ref{sec:Vendor Reports Analysis}. Then in Section~\ref{sec:carbon_cost_of_peripheral_components}, we compare the ACT tool with the PAIA tool, a commercial LCA tool to show that the SOTA carbon estimation tool doesn't consider the peripheral parts of a real-world computing system.
\vspace{-5pt}
\subsection{Vendor Reports Analysis}
\vspace{-5pt}
\label{sec:Vendor Reports Analysis}
Current server vendors, including HP, Dell, and Lenovo, the largest server vendors in the global server market, provide carbon footprint reports for their server products. 
We collect all the available carbon footprint reports, in total 96 different reports from their corresponding official websites~\cite{Dell-ENV,HP-ENV,Lenovo-ENV}. 
These reports contain servers released ranging from 2014 to 2022. 
All of the 3 vendors, i.e., Dell, HP, and Lenovo, leverage the PAIA LCA tool flow. 
The carbon cost is reported as green house gas emissions in  \SI{}{\kilo \gram \coeq}.

The life cycle of hardware products can be divided into 4 consecutive phases: manufacturing, transportation, operation use, and end-of-life (EOL) processing
Most of the reports give the carbon cost for each phase.
Generally, in a 4-year or a 5-year lifetime as reported, the operational cost takes a significant percentage, which consumes 70.2\% to 90.2\% 
in Dell's reports, 66.2\% to 93.7\%, in HP and  39.4\% to 97.0\% Lenovo's reports. 
In these reports, the manufacturing phase also takes a considerable proportion of the carbon cost, which consumes 9.5\% to 27.7\% for Dell, 5.4\% to 28.4\% for HP, and 2.9\% to 59.5\% for Lenovo. 
The transportation and EOL carbon costs are relatively small in the four phases and usually take less than 1\% of all the costs.
\begin{table}
\centering
\small
\tabcolsep 3pt
\caption{
Comparison among servers with the highest and lowest manufacturing/embodied carbon cost, i.e., green house gas emissions in \SI{}{\kilo \gram \coeq}, from the three vendors.
}
\vspace{-0.2cm}
\label{tab:embodied_carbon_breakdown}
\begin{adjustbox}{width=1\columnwidth,center}
\begin{tabular}{c|c|c|c|c|c|c}
\hline
\textbf{Components}
& \textbf{Dell-H}
& \textbf{Dell-L}
& \textbf{HP-H}
& \textbf{HP-L}
& \textbf{Lenovo-H}
& \textbf{Lenovo-L}
\\ \hline
Total embodied & 1782 & 1133 & 3880 & 423 & 15593 & 585\\ \hline
HDD carbon & N/A & N/A & N/A & 0 & 0 & 0\\ \hline
SSD carbon & N/A & N/A & N/A & 18 & 85 & 30\\ \hline
Mainboard carbon & N/A & N/A & N/A & 203 & 15167 & 426\\ \hline
Daughterboard carbon & N/A & N/A & N/A & 128 & 256 & 34\\ \hline
Others (PSUs, fans, etc.) & N/A & N/A & N/A & 74 & 85 & 95\\ \hline \hline
Server Name& r930 & t340 & DL380 & DL20 & SR950 & SR250v2\\\hline
\end{tabular}
\end{adjustbox}
\vspace{-20pt}
\end{table}

\textbf{Breakdown of the manufacturing/embodied cost:} 
The manufacturing or embodied carbon cost of a server contains the carbon cost from different parts and the assembly cost. 
The main parts of a server contain the parts with integrated circuits (IC), i.e., mainboard, daughterboard, memory, SSD, and HDD, and non-IC parts, i.e., chassis, power supply units (PSUs), and fans.
Among the three vendors, only Lenovo gives the detailed manufacturing or embodied carbon cost breakdown of all the servers.
HP includes the breakdown in reports of some servers while Dell does not provide such information at all. 
We show the manufacturing or embodied carbon cost breakdown for servers with the highest and lowest embodied carbon cost from the three vendors in Table~\ref{tab:embodied_carbon_breakdown}.
As reported by Lenovo, the IC parts take the largest proportion, 84\% to 99.5\% of the embodied cost. 
Besides, the non-IC parts take 0.5\% to 16\% of the overall embodied carbon.

\textbf{Server Configurations Breakdown}
\label{sec:observation_why_hp}
The detailed server configurations used in reporting the carbon cost are important to understand where the embodied carbon comes from.
We list the reported server configurations from the three vendors in Table~\ref{tab:config_breakdown}.
For example, when reporting manufacturing/embodied carbon cost, the reported configurations include number of CPUs (\#CPUs), total number of CPU cores, main memory size in GB, number of SSD/HDD disks, total SSD/HDD storage size (GB), number of PSUs, fans, server types (rack, blade, tower). 
We perform feature importance analysis and find that \emph{among all configurations, the total number of CPU cores, main memory size,  total SSD/HDD storage size, and release year, are features with the highest importance scores in estimating the manufacturing/embodied carbon cost.} 
This is intuitive to understand. 
For example, the total number of CPU cores correlates with the CPU chip die area and the release year, which implicitly indicates the fabrication technology node. They together determines the total carbon cost of the CPUs.

\begin{table}
\centering
\small
\caption{
Server configurations used for the carbon footprint report across vendors. A star~\mystar ~ means the configuration is ``important" in estimating the manufacturing/embodied cost.
}
\vspace{-8pt}
\label{tab:config_breakdown}
\begin{adjustbox}{width=0.8\columnwidth,center}
\begin{tabular}{c|c|c|c|c}
\hline
&  & HP & Dell & Lenovo \\ \hline 
\multirow{10}{*}{{Manufacturing}} & \#CPUs & \checkmark & \checkmark & \checkmark \\ 
& Total CPU \#Cores~\mystar  & \checkmark & & \\ 
& Memory (GB)~\mystar  & \checkmark & \checkmark & \\ 
& SSD \#disks  & \checkmark & \checkmark & \checkmark \\ 
& Total SSD size (GB)~\mystar  & \checkmark & & \\ 
& HDD \#disks  & \checkmark & \checkmark & \checkmark \\ 
& \#Fans &&& \checkmark \\
& \#PSU & \checkmark && \\
& Server type & \checkmark & \checkmark & \checkmark \\ 
& Release year~\mystar  & \checkmark & \checkmark & \checkmark \\ \hline
\multirow{2}{*}{{Transport}} & Region & \checkmark & \checkmark & \checkmark \\
& Server Weight & \checkmark & \checkmark & \checkmark \\
& Transportation Method & && \checkmark \\ \hline
\multirow{3}{*}{{Use}} & Product Lifetime & \checkmark & \checkmark & \checkmark \\
& Yearly Energy (TEC) & \checkmark & \checkmark & \\
& TEC Configs & \checkmark && \\ \hline
EOL & EOL Methods & && \checkmark \\ \hline
\end{tabular}
\end{adjustbox}
\vspace{-10pt}
\end{table}

Among all the three vendors, HP reports all the ``important" server configurations in the carbon footprint reports, which allows us to use the data in these reports to build an embodied carbon cost model and to predict the embodied carbon cost of a certain server if given these server setup features as the input.

\vspace{-5pt}
\subsection{Comparsion between ACT and PAIA tool}
\label{sec:carbon_cost_of_peripheral_components}
\vspace{-5pt}

\begin{figure}[tb]
\centering
\includegraphics[width=1\columnwidth]{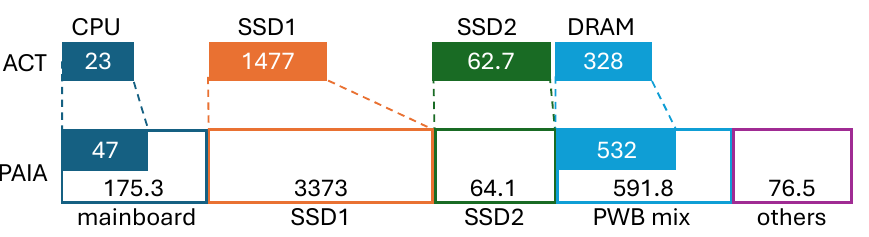}
\caption{The comparison between reports of ACT and PAIA for the same Dell R740 server setup. Carbon cost in the figure are using \SI{}{\kilo\gram\coeq} as unit.}
\label{fig:report_comparasion}
\vspace{-20pt}
\end{figure}

By comparing and analyzing the two reported carbon cost from ACT~\cite{ACT} and PAIA~\cite{LCA_R740} for the Dell R740 server, 
we show that the peripheral devices contribute a considerable proportion to the overall carbon cost, which the SOTA approaches like ACT omitted to discuss.

The Dell R740 rack server is equipped with a mainboard installed with 2x Intel Xeon 6152 CPUs, 8x 3.84TB SSDs (SSD1), 1x 400GB SSD (SSD2), and 12x 32GB DRAM. 
The server also contains 3 riser cards (devices for adding more expansion cards to the mother board, such as the PCIE slot), an HDD controller, an ethernet, and an Ethernet controller, as well as PSUs, fans, and chassis.

As shown in Figure \ref{fig:report_comparasion}, we break down the reports from the two tools. 
The part with the largest gap is the \textbf{CPU vs. mainboard} pair, where a \textbf{7.6x} gap exists between the two reports (23.14 vs. 175.3). 
In these parts, for the CPU alone, a roughly 2x gap exists, and another 5.6x carbon cost comes from the other part of the mainboard which ACT omitted to discuss, including the mainboard printed wiring board (PWB)  (108.68 \SI{}{\kilo\gram\coeq}), connectors (7.36 \SI{}{\kilo\gram\coeq}), and the transportation in the manufacturing phase (12.80 \SI{}{\kilo\gram\coeq}).

For the storage and memory devices including SSDs and DRAMs, there exists a 2.3x and 1.61x gap for SSD1 and DRAM, For the SSD2, the two tools made a similar estimation. 

ACT also ignored the riser cards, HDD controller, ethernet, and ethernet controller, which are reported by Dell in the \textbf{PWB mix} category, as well as the PSU, chassis, and fans, which are represented in the \textbf{others} category. 
These components take 136.3 \SI{}{\kilo\gram\coeq}, reaching 77\% of the mainboard's carbon cost.

In total, there is a 2.26x 
gap between ACT's and PAIA's reports. 
To be noted, in this server setup, there are in total 32\SI{}{\tera\byte} of SSDs, taking most of embodied carbon cost. 
In other GPU servers with smaller SSD and DRAM sizes, the gap between the ACT tool and the PAIA tool can be larger. 
If we use the setup of Dell R740 in Dell's carbon footprint report~\cite{Dell_r740_carbon_footprint}, the gap between the 2 tools can be 6.81x. 
In this setup, the server has only 1.6TB HDD and 32GB DRAM.

Such a huge gap can not be solved by traditional bottom-up methods since the peripheral components are usually not reported by the vendors.
In Section~\ref{sec:modeling}, we show that by applying an end-to-end method, the SCARIF framework can mitigate these gaps.

\section{SCARIF Modeling and Framework}
\vspace{-5pt}
\label{sec:modeling}
In this section, we present SCARIF modeling and framework. 
We first illustrate how we build the SCARIF modeling in estimating the embodied carbon cost given a certain server setup, and how we choose the parameters in the modeling in Section~\ref{sec:model_build}.
Then we show how to use SCARIF modeling to predict the embodied carbon cost for servers across different vendors in Section~\ref{sec:model_valid}.
We integrate modeling for the accelerator embodied carbon cost into the SCARIF framework by utilizing a modified version of the ACT tool in Section~\ref{sec:model_accelerator}.
\vspace{-5pt}
\subsection{SCARIF Tool Overview}
\vspace{-5pt}
\label{sec:model_build}

\begin{figure}[tb]
\centering
\vspace{-10pt}
\includegraphics[width=1\columnwidth]{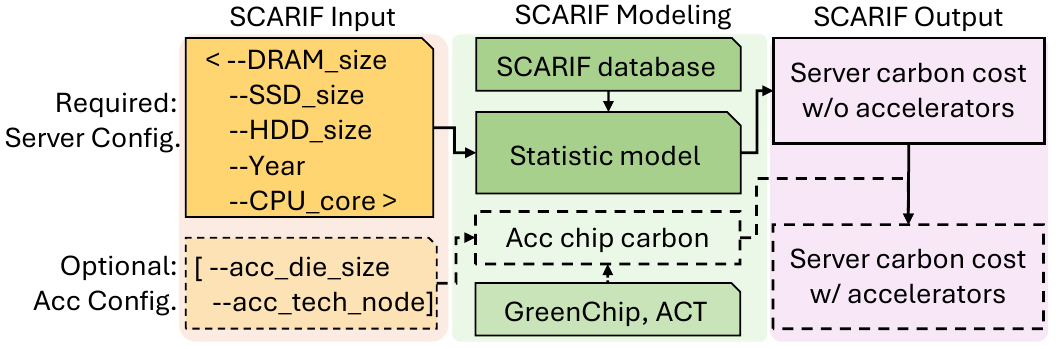}
\caption{The overview of SCARIF tool.}
\label{fig_io}
\vspace{-20pt}
\end{figure}

As analyzed in Section~\ref{sec:observation_why_hp}, we choose to use the reports from HP to build up the model for embodied cost estimation.

Currently, 9 reports from HP have the full details of the total number of CPU cores, SSD size, and memory size. 
We chose to use a linear model in the modeling by using these data. 
In the following sections, we further discuss the parameters used in the modeling in detail.
The input and output of SCARIF tool are as shown in~\ref{fig_io}.

\textbf{Total number of CPU cores:} 
The embodied cost of CPUs, one of the largest parts of the overall embodied cost, is proportional to the CPU chip die area when the other conditions are fixed. 
In other words, when we consider the CPUs manufactured using the same technology node, the yield is considered fixed, and the embodied cost $E_{CPU}$ can be characterized as
$E_{CPU}=K_1\cdot Area_{CPU}$
where $K_1$ is the learnable coefficient and $Area_{CPU}$ is the CPU chip die area.

However, the specified part names of CPUs are quite rare in current reports, making it impossible to get the actual chip area, thus we have to use an alternative to estimate the carbon cost of CPUs. 
Here we choose the total number of CPU cores, which is the multiplication of the number of CPUs in a server and the maximum number (if the exact number is not reported) of cores per CPU a server can support. 
This parameter is positively correlated to the chip area.

Moreover, the peripheral components of the server, such as the mainboard chipsets, usually share the same trend of the number of CPU cores. 
For example, the motherboard chipset assists the CPU to communicate with the peripheral components like memory and disk. 
The Intel chipset C236 used by the Dell R230 server, which can contain at most 1 CPU/4 total CPU cores, has a package area of 529 $mm^2$ with a TDP of 6W. 
Whereas the Intel chipset 602 used by the Dell R930 server, which can contain at most 4 CPUs/96 total CPU cores, has a package area of 729 $mm^2$ with a TDP of 8W. 

\textbf{SSD, HDD, and memory size:} The carbon cost of storage devices, like SSD and HDD, can be quantified using their sizes when the technology nodes are the same.
Similarly, the carbon cost of memory can be quantified using the memory size when the DDR technology is given.
Therefore, the size of SSD, HDD, and memory can be used in the modeling. 

\textbf{Year:} The discussions above are based on the assumption of using the same technology node. 
However, a new technology node appears and will be applied to new chips every few years. 
To represent the effect of different technology nodes, we consider the release year of the server, as a representation for different technology nodes. 
To emphasize the changes in the recent decades, we use $(Year-2000)$ as the parameters.

Therefore, the overall modeling is as follows:
\vspace{-5pt}
\begin{equation}
    \begin{aligned}
   E=K_1\cdot \#CPUcores +K_2\cdot Size_{SSD}
    +K_3 \\ \cdot Size_{HDD}+ K_4\cdot Size_{Mem}
    +K_5\cdot (year-2000) + D
    \end{aligned}
\end{equation}
Where the $K_1$ to $K_5$ and $D$ are trainable parameters.

We choose to use the reported coefficient for HDD and SSD from prior work reported in the literature~\cite{dirty_ssd}:
\begin{equation}
    \begin{aligned}
   K_2= 0.16~KgCO_2e/GB, 
   K_3 = 0.04~KgCO_2e/GB
    \end{aligned}
\end{equation}

The other coefficients learned from the HP dataset are:
\vspace{-3pt}
\begin{equation}
    \begin{aligned}
   K_1=5.01~KgCO_2e, 
   K_4=0.95~KgCO_2e/GB\\
   K_5=83.08~KgCO_2e/year\\
    \end{aligned}
\end{equation}

To improve the model's flexibility, the intercept $D$ is set as a hyperparameter. On the HP dataset, the $D$ is set to -1100.

\vspace{-3pt}
\subsection{Modeling Validation Across Vendors}
\label{sec:model_valid}
\vspace{-3pt}

To demonstrate the effectiveness of our modeling, we apply our model to the data reported by other vendors: Dell and HP. 
Due to the missing configuration information in these reports, some augmentation methods to the data are applied.
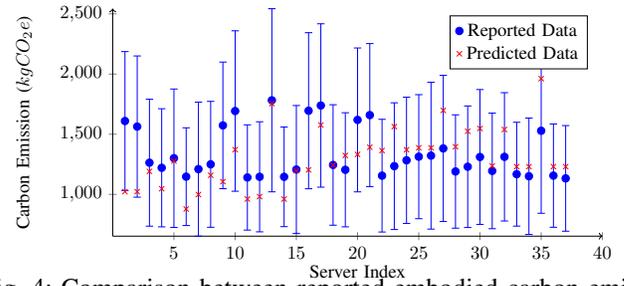
\begin{figure}[tbp]
  \centering
  \pgfplotstableread{
x         y    y-min  y-max
1	1609.65	577.22	577.22
2	1563.54	586.96	586.96
3	1263.24	527.57	527.57
4	1220.87	490.96	490.96
5	1300.65	575.36	575.36
6	1147.20	406.30	406.30
7	1209.84	557.21	557.21
8	1249.76	523.26	523.26
9	1573.36	526.30	526.30
10	1692.75	666.74	666.74
11	1140.56	437.57	437.57
12	1146.08	457.60	457.60
13	1782.20	761.12	761.12
14	1145.94	414.26	414.26
15	1206.50	532.00	532.00
16	1695.33	648.65	648.65
17	1738.80	678.96	678.96
18	1244.40	500.69	500.69
19	1204.35	474.34	474.34
20	1618.82	598.05	598.05
21	1659.00	594.40	594.40
22	1155.52	468.49	468.49
23	1234.50	526.80	526.80
24	1283.18	524.56	524.56
25	1313.28	515.58	515.58
26	1321.92	610.56	610.56
27	1381.80	607.60	607.60
28	1190.16	471.19	471.19
29	1229.68	505.25	505.25
30	1310.80	560.48	560.48
31	1194.93	480.42	480.42
32	1312.00	533.33	533.33
33	1167.72	431.57	431.57
34	1150.60	484.00	484.00
35	1528.80	686.00	686.00
36	1155.84	429.31	429.31
37	1132.86	438.92	438.92
}{\reporteddata}

\pgfplotstableread{
x         y    
1 1022.8
2 1022.8
3 1190.8
4 1046.8
5 1278
6 878.4
7 998.6
8 1158.8
9 1105.8
10 1371.9
11 961.4
12 981.5
13 1754.4
14 961.4
15 1196.9
16 1204.1
17 1576.9
18 1243.2
19 1323.3
20 1332.1
21 1392.1
22 1364
23 1563.2
24 1371.3
25 1387.3
26 1387.3
27 1697.7
28 1396.2
29 1524.6
30 1547.4
31 1235.9
32 1539.3
33 1230.7
34 1230.7
35 1962.2
36 1230.7
37 1230.7
}{\predictdata}

\begin{tikzpicture}[scale=0.7] 
\begin{axis}[
    axis lines=middle,
    xmin=0, xmax=40,
    axis line style={->},
    x label style={at={(axis description cs:0.5,-0.1)},anchor=north},
    xtick distance=5,
    y label style={at={(axis description cs:-0.15,.5)},rotate=90,anchor=south},
    xlabel={Server Index},
    ylabel={Carbon Emission ($kgCO_2e$)},
    legend pos=north east,
    width=.6\textwidth,
    height=.25\textheight]
    \addplot[blue, only marks] 
        plot[error bars/.cd, y dir=both, y explicit]
        table[x=x,y=y,y error plus expr=\thisrow{y-max},y error minus expr=\thisrow{y-min}] {\reporteddata};
        
    \addplot[red, only marks,mark=x] 
  plot[error bars/.cd, y dir=both, y explicit]
  table[x=x,y=y] {\predictdata};
  \legend{Reported Data, Predicted Data}
\end{axis} 
\end{tikzpicture}
  \vspace{-10pt}
  \caption{Comparison between reported embodied carbon emission and predicted embodied carbon emission on 37 servers reported by Dell.}
  \label{fig:prediction}
  \vspace{-20pt}
\end{figure}

\textbf{Applying SCARIF to Dell servers:} 
Dell reports the size of memory, HDD, and SSD, but does not report the number of CPUs nor the cores per CPU. 
To get a reasonable estimation for configurations on different Dell servers, we refer to the corresponding Dell server specification sheet to get the maximum number of CPU cores that are compatible with the specified server.
To represent the other effects across vendors, we add another constant $D'$ = -400 to the interception. 
As shown in Figure~\ref{fig:prediction}, the circled data with error bars represent the reported data from Dell.
The error bars come from the fact that Dell reports the standard deviation for the embodied carbon cost.
The cross points represent the data predicted from our SCARIF modeling.  
We show the error bars with 0.4 of the standard deviation, $\sigma$, and our modeling data is always within $0.4\times\sigma$ of the Dell reported data, producing an acceptable estimation result. 
The average error is within $0.15\times\sigma$.

\textbf{Applying SCARIF to Lenovo servers:} 
We also apply our modeling and compare it with the Lenovo data (in total, 40 reports). 
Lenovo does not report the CPU part name, either.
We use the same method, i.e., referring to the server specification sheets, to estimate the CPU core number. 
Similarly, we add another constant $D''$ = -900 to the interception for Lenovo in the modeling to represent the effects across vendors.
The average error of our modeling on Lenovo data is 50\%.
The errors come from: (1) Lenovo does not report memory size and we use 64GB in our modeling; (2) Lenovo has 13 server reports with relatively large embodied carbon cost, that is, the Lenovo reported carbon is over 2x of our modeling results.

\vspace{-10pt}
\subsection{Embodied Carbon Cost Modeling for Accelerators}
\label{sec:model_accelerator}
\vspace{-5pt}

Hardware accelerators such as GPUs and FPGAs are important in today's computing systems, particularly in modern datacenters. 
However, none of the vendors provide the carbon footprint report data for servers equipped with accelerators, 
and the GPU and FPGA vendors are not as forthcoming with sustainability data.
Here, we propose a method to roughly estimate the carbon cost of such systems by utilizing the ACT tool and some scaling assumptions of accelerators.

Different from CPUs, there are fewer models of accelerators but more varieties of architectures among the models. 
As a result, a unified feature like CPU core numbers is impractical for accelerators since the GPU core usually takes a different area from the CPU core, and there is no real core equivalent in an FPGA. 
Here, we assume that the GPU-related part, which contains not only the chip itself but also the peripheral components like the slots, shares the same pattern as the accelerator-related part. 
Thus, the carbon of the accelerator-related part can be obtained by analyzing the carbon of the CPU-related part and the chip area of both parts. 
We leverage the ACT tool in generating the estimation. 
Given the model of the accelerator, its carbon cost of the chip, $Chip(Acc)$ can be obtained by ACT. Then the accelerator part carbon cost $System(Acc)$ can be roughly estimated as:
\begin{equation}
\frac{System(Acc)}{Chip(Acc)} = \frac{System(CPU)}{Chip(CPU)} = K_6
\end{equation}
Where $K_6$ is a hyper-parameter which can be obtained by integrating different CPU-only server setups as the baseline. 

For example, for a Dell's R740 server with 2 Intel Xeon 8180 CPUs, 1TB HDD, 64GB memory, and an Nvidia V100 GPU, the ACT tool reports the embodied costs of
26.71 and 15.69 
\SI{}{\kilo\gram\coeq}
for CPU and GPU, respectively. 
While SCARIF reports these costs as 
280.56, and 164.80 \SI{}{\kilo\gram\coeq}.
The embodied carbon cost of such a CPU server without the GPU accelerator is 1993.72 \SI{}{\kilo\gram\coeq} and 2158.53 \SI{}{\kilo\gram\coeq} with the GPU accelerator. 
The embodied carbon cost estimation for FPGAs in our modeling is similar to that of GPUs. 
We provide more detailed carbon analysis on servers with accelerators in Section~\ref{sec:analysis}. 
\section{Carbon Analysis From SCARIF}
\vspace{-5pt}
\label{sec:analysis}
In this section, we show how can we holisticly view the server carbon footprints by using SCARIF for several different scenarios under given workloads.
We provide concrete case studies and analysis: (1) to compare both manufacturing and operational costs for systems; (2) to demonstrate the breakeven analysis when upgrading servers; (3) to compare system with different kinds of accelerators, i.e., GPUs or FPGAs.
\vspace{-5pt}
\subsection{Analysis and Experimental Setup}
\vspace{-5pt}
\textbf{Hardware configurations.} 
We use two generations of servers in our analysis, Dell R740 and Dell R750, released in 2017 and 2020 respectively.
For Dell R740, we configure 2x Intel 14nm Xeon 8180 CPUs with 694\SI{}{\milli\meter\squared} chip area, 64GB DDR4-10nm DRAM, and 1TB HDD. 
For Dell R750, we configure 2x Intel 10nm Xeon 8375 CPUs with 660\SI{}{\milli\meter\squared} chip area, 64GB DDR4-10nm DRAM, and 1TB HDD. 
For the accelerators, we choose NVIDIA V100 GPU and NVIDIA A100 GPU, the 12nm V100 was released in 2017, with a chip area of 815 \SI{}{\milli\meter\squared}, whereas the 7nm A100 was released in 2020, with a chip area of 826 \SI{}{\milli\meter\squared}. We also choose the Xilinx ZCU102 FPGA for comparison, which was released in 2016, has a fabrication of 16nm, and a chip area of 245 \SI{}{\milli\meter\squared}.

\begin{table}
\small
\tabcolsep 3pt
\caption{
Performance and power comparisons among Intel CPUs Xeon 8180 and 8375, Nvidia GPUs V100 and A100, and AMD FPGA ZCU102.
}
\vspace{-8pt}
\label{tab:performance}
\begin{adjustbox}{width=1\columnwidth,center}
\begin{tabular}{c|c|c|c|c|c}
\hline
& Xeon 8180 & Xeon 8375 & V100 & A100 & ZCU102 \\ \hline
Latency (ms) & 217.98 & 176.68 & 2.96 & 1.84 & 32.72 \\ \hline
Power (W) & 205 & 300 & 250 & 175 & 25  \\ \hline
Static Power (W) & 10 & 10 & 39 & 53 & 1
\\ \hline
Framework & ONNX & ONNX & TensorRT & TensorRT & HeatViT \\ \hline
\end{tabular}
\end{adjustbox}
\vspace{-20pt}
\end{table}

\textbf{Application workload characterization.} We choose a representative deep learning workload, the inference of a vision transformer model DeiT-T \cite{touvron2021training_deit_t}.  
We measure the performance in latency and power on different hardware including using CPU only, using GPU, and FPGA as shown in Table~\ref{tab:performance}. 
When running on CPUs, we use onnx and onnx\_runtime \cite{onnx}. To be noted,the CPU performance in Table~\ref{tab:performance} is reported on using one CPU core, and the power is reported on using all available 28 CPU cores for Xeon 8180 or 32 CPU cores for Xeon 8375. We will use these data accordingly to normalize the power, utilization, etc. 
When running on accelerators, we use the state-of-the-art GPU framework and FPGA backends, that is, TensorRT~\cite{TensorRT} for GPU, and HeatViT~\cite{dong2023heatvit} for FPGA.

The GPU or FPGA accelerator card is hosted in the CPU server. For simplicity, we denote the Dell R740 server mentioned above as server 1, and the R750 server as server 2. In the following sections' discussion, we have five different types of system setups (DRAM, disk information omitted for brevity): 
(i) \texttt{System 1}:  Server 1 with CPUs and GPU V100; (ii) \texttt{System 2}: Server 2 with CPUs and GPU A100; (iii) \texttt{System 3}: Server 1 with CPUs and FPGA ZCU102.

When we compare the total carbon cost of the two systems, we calculate the embodied carbon for each system and the operational carbon cost when running certain workloads on such systems for certain periods of lifetimes (years) with certain accelerator utilization. 
We also normalize the number of servers or normalize the accelerator utilization when comparing two systems based on the assumption that the total workloads/tasks are the same.

When calculating the operational carbon cost, we use the carbon intensity results from ~\cite{ollivier2022sustainable} to convert the annual energy consumption to the greenhouse gas emission. That is, for 4 different states in the US: Arizona (AZ), California (CA), Texas (TX), and New York (NY), the carbon intensities (in \SI{}{{\kilo\gram\coeq}\per{\kilo\watt\hour}}) are 0.395, 0.234, 0.438, and 0.188.
\vspace{-5pt}
\subsection{Tradeoff in Upgrading Servers}
\vspace{-5pt}

Servers with a newer technology node usually have better energy efficiency in the operation phase, but upgrading new servers requires an instant embodied carbon cost. To demonstrate the tradeoff,
we use \texttt{System 1} (2017 Servers with 2017 GPU) and \texttt{System 2} (2020 Servers with 2020 GPU).
For the embodied cost, the R740+1 V100 server takes 2159 \SI{}{\kilo\gram\coeq} in total, and the R750+1 A100 server takes 2542 \SI{}{\kilo\gram\coeq}. 
For the operational cost, we assume \texttt{System 1} running at 100\% accelerator utilization, and the accelerator utilization of \texttt{System 2} is 62.2\%.
After normalization, the estimated annual energy consumption of \texttt{System 1} and \texttt{System 2} is 2365 kWh and 1304 kWh, respectively. 
Due to the different assumptions of the grid mix in four states, the annual operational carbon cost will be different.

\begin{figure}[tb]
\centering
\includegraphics[width=0.9\columnwidth]{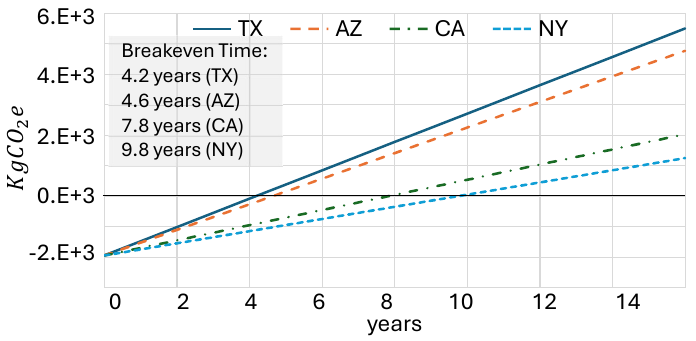}
\vspace{-5pt}
\caption{Carbon savings of upgrading vs. non-upgrading servers in four different states in United States.}
\label{fig:GPU_upgrade}
\vspace{-15pt}
\end{figure}

The carbon saving of upgrading to R750 compared with non-upgrading is as shown in Figure~\ref{fig:GPU_upgrade}, where the 4 different lines stand for different carbon intensity assumptions in four states. 
Initially, the carbon costs of 4 lines start at the same point (year=0), with the negative embodied carbon cost as the new servers require embodied cost to be paid instantly once the server is upgraded. 
Then all 4 lines go up since the new server has a lower energy consumption in the operational phase. 
Based on different carbon intensity assumptions, i.e., in different regions, 
it will take 4.2, 4.6, 7.8, and 9.8 years to reach the breakeven point, where two strategies (non-upgrading vs upgrading) have the same overall carbon cost. 
Such a long break-even time is longer than the estimated 4-year server lifetime from the vendors. In other words, upgrading these servers will never have a gain in carbon efficiency. 
Instead, when using ACT to estimate carbon emissions, the embodied carbon cost of \texttt{System 2} will be about 70 \SI{}{\kilo\gram\coeq}, leading to a break-even time of less than 1 year for all four locations.
The design choice will be flipped, which calls for the use of SCARIF, which characterizes more accurately on the system level instead of device level carbon cost.

\vspace{-5pt}
\subsection{Carbon Cost for Systems with GPU and FPGA}
\vspace{-5pt}


\begin{figure}[tb]
\centering
\includegraphics[width=1\columnwidth]{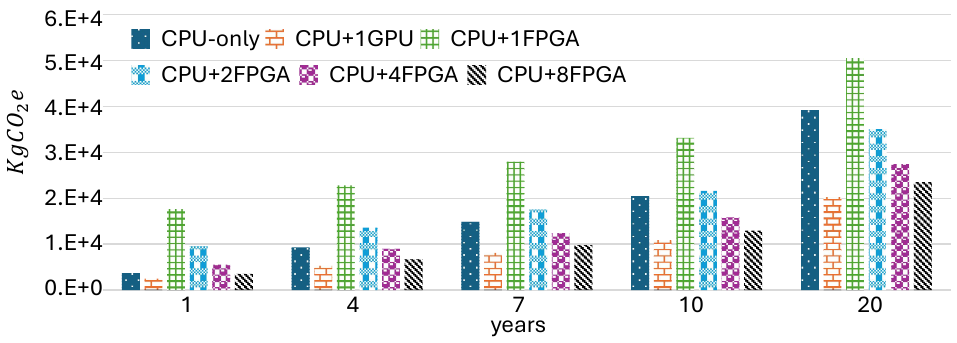}
\vspace{-15pt}
\caption{Overall carbon cost comparison among CPU-only, 1-GPU and 1-,2-,4-,8-FPGA servers. We assume accelerators are all running at 100\% utilization.}
\label{fig:GPU_vs_FPGA}
\vspace{-10pt}
\end{figure}

We use \texttt{System 1} (2017 Servers with 2017 GPU) and \texttt{System 3} (2017 Servers with 2016 FPGA).
We assume all the systems with accelerators running at 100\% utilization with scaled numbers of servers. 
The result of the total carbon cost of the 6 configurations (CPU-only, CPU+1GPU, CPU with 1/2/4/8 FPGAs) is shown in Figure~\ref{fig:GPU_vs_FPGA}. 
\noindent\textbf{The system with 8-FPGA servers has lower carbon cost than with 1-FPGA servers, however, higher than that with 1-GPU servers}.
8-FPGA server systems have lower total carbon costs than 1-FPGA server systems because of lower waste on CPU (embodied and operational) carbon costs.
However, in terms of task efficiency per carbon cost, the GPU system is better than the FPGA system, which makes the 1-GPU server system have a lower carbon cost than the 8-FPGA server. 
\vspace{-5pt}
\section{Conclusion and Acknowledge}
SCARIF provides a key tool for modeling the embodied carbon cost of server systems with accelerators. We will add SCARIF with more features to further guide the system carbon emission optimization from edge-level to enterprise-level.

We acknowledge NSF support (\#2213701, \#2217003, \#2324864, \#2328972) and the University of Pittsburgh New Faculty Start-up Grant. We thank the AMD/Xilinx University Program for hardware and software donations.

\bibliographystyle{IEEEtran}
\bibliography{references}

\end{document}